\documentclass[amsmath,amssymb,aps,floatfix,prd,a4paper]{revtex4}
\usepackage{epsfig}%
\usepackage{amsmath}%
\usepackage{graphicx}
\usepackage{amsmath}
\usepackage{amsfonts}
\usepackage{amssymb}
\usepackage{physics}
\usepackage{changepage}
\usepackage{caption}
\usepackage{verbatim}
\usepackage{multirow}
\usepackage{color}
\usepackage{cancel}
\newcommand{\beq}{\begin{equation}}
\newcommand{\eeq}{\end{equation}}

\graphicspath{{Figs/}}
\begin{document}
\title{Emergence of KNO scaling in multiplicity distributions in jets produced at the LHC}
\author{G. R.  Germano}
\email[e-mail: ]{ guilherme.germano@usp.br}
\author{F. S. Navarra} 
\email[e-mail: ]{ navarra@if.usp.br} 
\affiliation{Instituto de F\'{\i}sica, Universidade de S\~{a}o Paulo, 
  Rua do Mat\~ao, 1371, CEP 05508-090, Cidade Universit\'aria, 
  S\~{a}o Paulo, SP, Brazil.}
\author{G. Wilk} 
\email[e-mail: ]{grzegorz.wilk@ncbj.gov.pl}
\affiliation{National Centre For Nuclear Research, Pasteura 7, Warsaw 02-093, Poland.}
\author{Z. Wlodarczyk}
\email[e-mail: ]{zbigniew.wlodarczyk@ujk.edu.pl}
\affiliation{Institute of Physics, Jan Kochanowski University, 25-406 Kielce, Poland.}

\begin{abstract}
In this work we study the multiplicity distributions (MDs) of charged particles within 
jets in proton-proton collisions, which were measured by the ATLAS 
collaboration in 2011, 2016 and 2019. The first data set refers to jets with 
smaller transverse momenta ($4 < p_T < 40$ GeV ) whereas the other two refer 
to higher $p_T$ jets ($0.1 < p_T < 2.5$ TeV).   We find that the lower $p_T$ set shows no sign of KNO scaling and that the higher 
$p_T$  sets gradually approach the scaling limit. For the lower $p_T$ set the mean multiplicity as a function of $p_T$ can be well 
described by expressions derived from QCD with different approximation schemes. For higher ($> 500$ GeV) values of $p_T$ these expressions 
significantly overshoot the data.  We show that the behavior of the MDs can be well represented by a Sub-Poisson 
distribution with energy dependent parameters. In  the range $40 < p_T < 100$ GeV 
there is a transition from sub to super poissonian behavior and the MD   
 evolves to a geometric distribution, which shows KNO scaling.  In this 
way we fit the MDs in all transverse momentum intervals with one single expression. 
We discuss the implications of this  phenomenological finding.

\end{abstract}
\maketitle

\section{Introduction}


Multiplicity is a global observable that allows to characterize
events in all colliding systems and has been widely studied in
attempts to understand multiparticle production processes. 
Experimentally, charged-particle multiplicity is one of the simplest  observables, 
and its importance stretches from calibration to advanced tagging techniques. 
We can try to obtain the maximum information from the multiplicity
distribution (MD) of charged particles to gain insights on the production   
mechanisms \cite{kittel,gor}. In high energy proton-proton collisions, 
particles  are produced basically in two ways. In an early stage of the collision there  
is a perturbative parton cascading process which is governed by the evolution   
equations of QCD. Later, the partons are converted into hadrons with additional 
particle production. Here the main mechanism is non-perturbative: string      
formation and decay.  The complete description of multiparticle production is 
very complicated  \cite{kittel,gor}. Nevertheless, in spite of the complexity 
of the subject, over the years the study of multiplicity distributions has   
given us valuable information about the dynamics of particle production. 

One of the remarkable features exhibited by MDs is the famous Koba-Nielsen-Olsen (KNO)
scaling \cite{kno,kno-1,kno-2,kno-3}, a phenomenon expected to be observed at asymptotically high energies. 
This prediction was made before the existence of QCD. Later there were several attempts to 
understand it in terms of quark and gluon dynamics, such as in Refs. \cite{dumi12,dumi13}.  In these works it was shown that 
KNO scaling emerges if the effective theory describing color charge fluctuations
at a scale of the order of the saturation momentum is approximately Gaussian. Moreover both  
non-linear saturation effects and running-coupling evolution are  required in order to obtain KNO scaling.
Very recently, in Ref.~\cite{kno1} a MD satisfying KNO scaling was derived by solving the Mueller dipole evolution 
equation in the double logarithm approximation. This supports the idea that gluon emission is a Markov process in 
which the emitted  gluons are strongly ordered in rapidity. 
 
From the experimental side there was progress too. The analysis of MDs in 
different systems and in different phase space regions showed that KNO scaling follows a 
complex pattern, appearing in certain situations and not in others. For example, in the 
analysis of minimum bias events in proton-proton collisions at $\sqrt{s} = 0.9, 2.36$ and  
$7$ TeV made by the CMS collaboration \cite{cms11},  KNO scaling appeared in the MD of 
particles in the central pseudo-rapidity region $|\eta| < 0.5$, whereas it was violated in 
the wider range $|\eta| < 2.4 $. More recently, violation of KNO scaling was also observed in studies of the
moments of the multiplicity distributions measured by ALICE and ATLAS data \cite{ge22,ku23} which were found 
to grow with the energy \cite{ge22,ku23}.

The higher energies reached at the LHC opened new ways to study   
multiparticle production.  Collimated groups of particles produced
by the hadronization of quarks and gluons are called jets. In hadron-hadron 
collisions, jets are produced in high-momentum transfer scatterings.
As the energy increases, we may produce jets with increasingly higher energies. 
These jets decay into more and more particles and they are now numerous enough 
so that we can study multiplicity distributions of particles produced in the jets. 
Multiplicity within jets is used to study both the perturbative
and non-perturbative QCD processes, and since quarks
and gluons have different color factors, the hadronization is
sensitive to the initial parton. Thus, the particle content and
its momentum distribution within jets can be used to discriminate
the type of parton that initiated the jet.
It is well known that gluon-initiated jets contain larger particle multiplicities
than quark-initiated jets at the same energy, and the
transverse momentum of the constituent particles is harder
for gluon-initiated jets \cite{opal98}.

The multiplicity distribution within low $p_T$  ($4 < p_T < 40$ GeV )  jets has been recently
addressed in \cite{greg22}, where the authors presented an analysis of the
ATLAS 2011 data  \cite{atlas11,dur}. They  showed 
that they can be well reproduced by a Sub-Poissonian (SP) distribution. This 
finding is interesting in itself since it establishes a clear difference between
the multiplicity distributions observed in minimum bias events and those observed 
in jets, the former being much broader than the latter. Triggering on high $p_T$ events, such as jets, 
one selects perturbative QCD processes. If the QCD parton radiation would be similar 
to bremsstrahlung, one would expect a multiplicity distribution similar to a 
Poisson distribution, which is much narrower than the familiar Negative 
Binomial Distribution (NBD), successfull in describing minimum bias data. Surprisingly,  
the appropriate distribution is SP, which is still narrower. All these 
considerations apply to the ATLAS data which refer to transverse momenta in the
range $4 < p_T < 40$ GeV.

In the theoretical analysis presented in  \cite{vertesi21,vertesi22}, 
the authors suggested that jet    
multiplicity distributions follow KNO scaling if one replaces the collision 
energy $\sqrt{s}$ by the jet average transverse momentum $p_T^{jet}$. To 
substantiate this conjecture the authors performed a simulation with the       
PYTHIA-8 Monte Carlo event generator. They obtained distributions that, when plotted 
in the KNO style, present a very good scaling.  Unfortunately, they missed the opportunity to compare 
the results of their simulations with the already existing data \cite{atlas11,atlas19}. 

Empirical scaling laws {\it per se} are important in physics, independently of their theoretical interpretation. 
To study them we first have to analyze the data choosing the most relevant variables and the best way to plot them. 
Then we fit these data with expressions which contain some physical meaning, such as, in the present context, the 
negative binomial distribution. The observation of scaling and the behavior of the fitting  distributions can give 
insights on the production dynamics and serve as a guide to  theoretical microscopic studies. 

In this work we will first revisit the ATLAS data and check whether they satisfy 
KNO scaling and also whether the average multiplicities are well described by  QCD predictions.  Then we will fit 
all the ATLAS  data with a Sub-Poisson distribution and analyze the energy dependence of the parameters.  As it will be seen, 
the data suggest that the MD undergoes a transition from sub to super poissonian behavior and starts to approach the KNO scaling 
limit.

\section{Revisiting the ATLAS data}

In this section we perform a quite simple and model independent exercise to check whether the existing ATLAS 
data \cite{atlas11,atlas19} satisfy  the scaling found in 
\cite{vertesi21,vertesi22}.  
\begin{figure}[!t]
\begin{tabular}{ccc}
\centerline{
{\includegraphics[height=4.5cm]{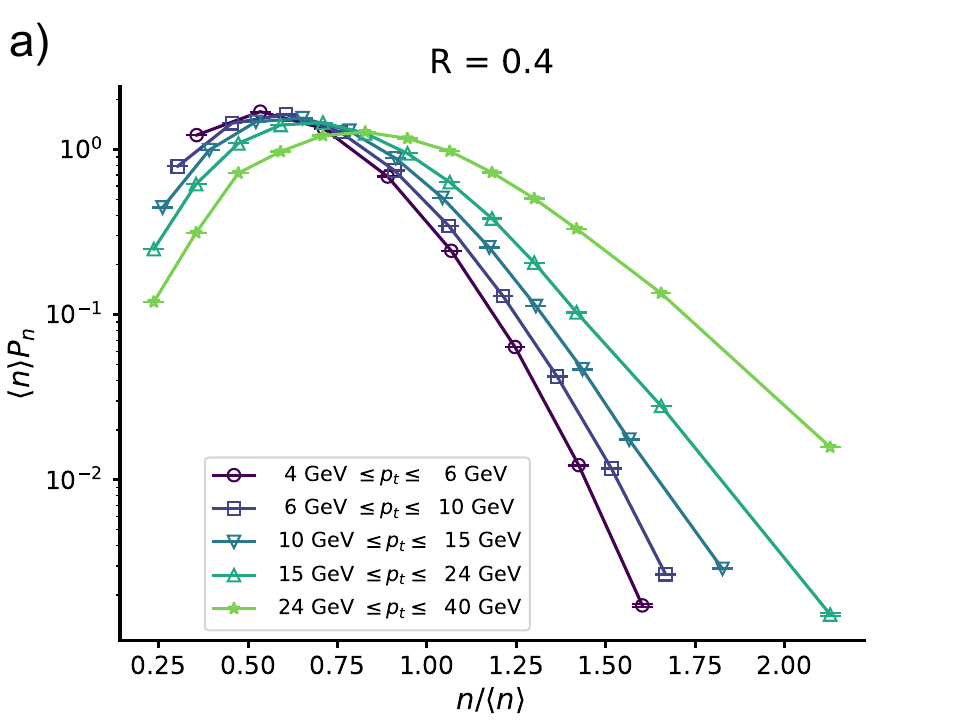}}
{\includegraphics[height=4.5cm]{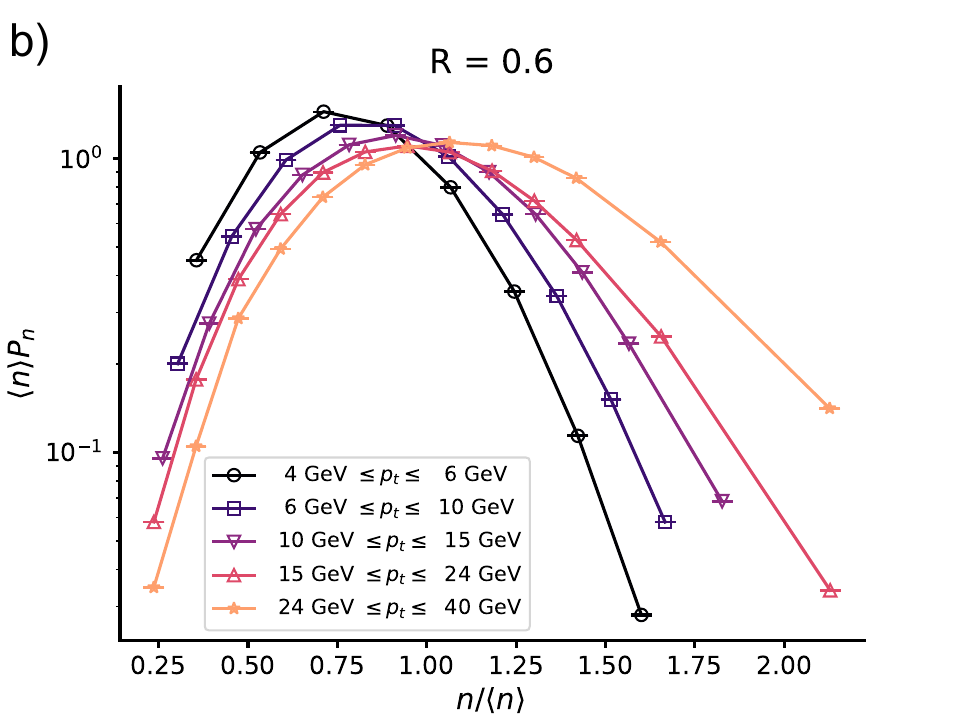}}
{\includegraphics[height=4.5cm]{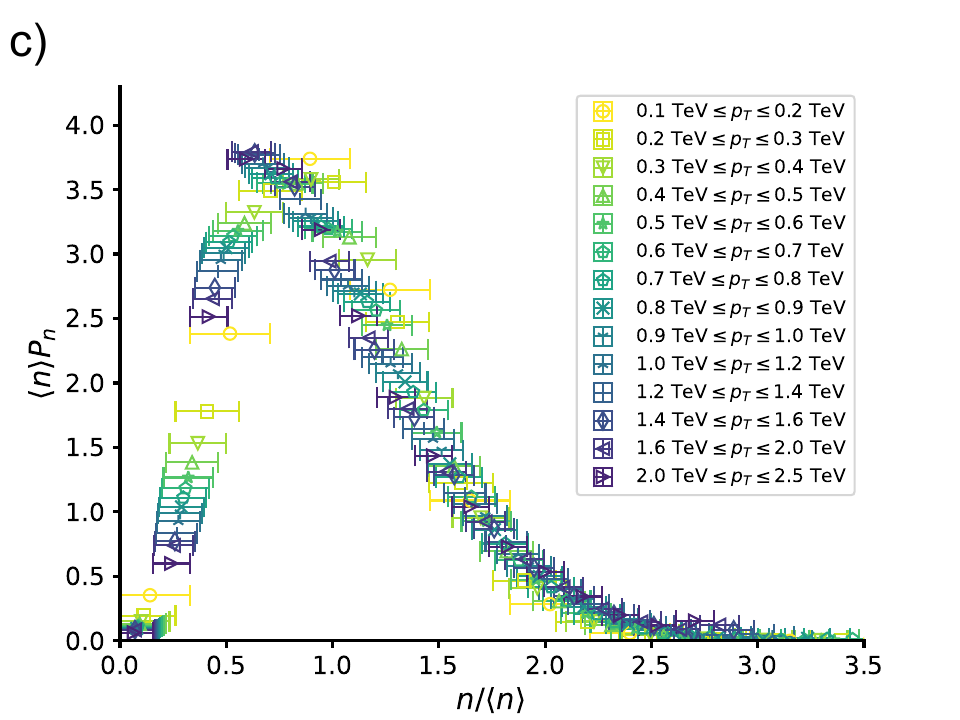}}
}
\end{tabular}
\caption{Data  from Ref.~\cite{atlas11} plotted in the KNO form
for  a) $R=0.4$, b)  $R=0.6$. c) Data from Ref.~\cite{atlas19}.}
\label{fig1}
\end{figure}
\begin{figure}[!t]
\begin{tabular}{cc}
\centerline{
{\includegraphics[height=5.0cm]{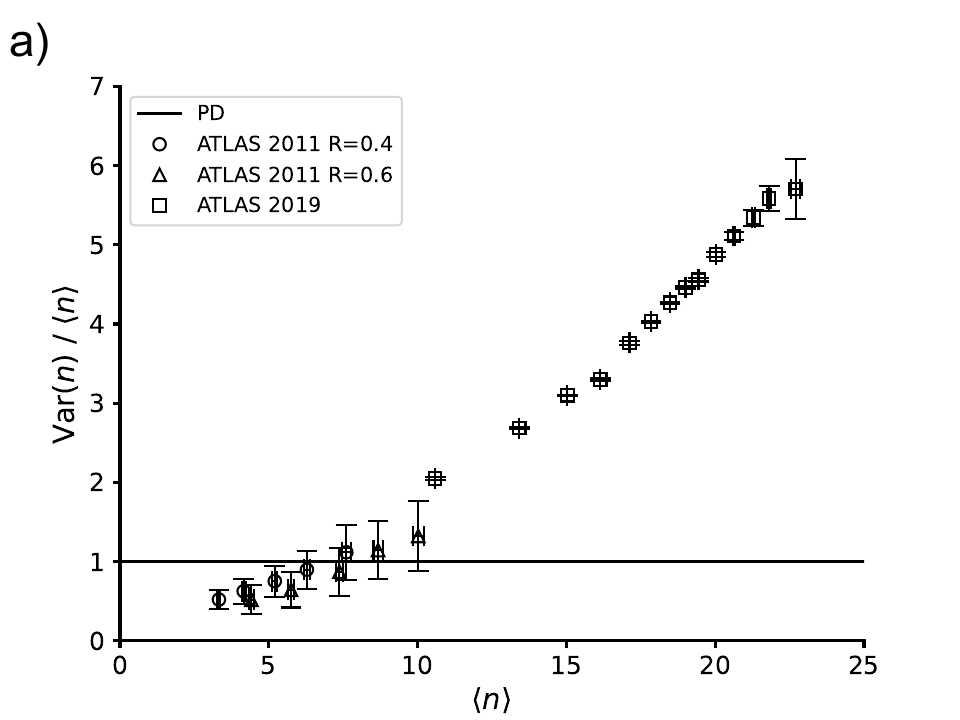}}
{\includegraphics[height=5.0cm]{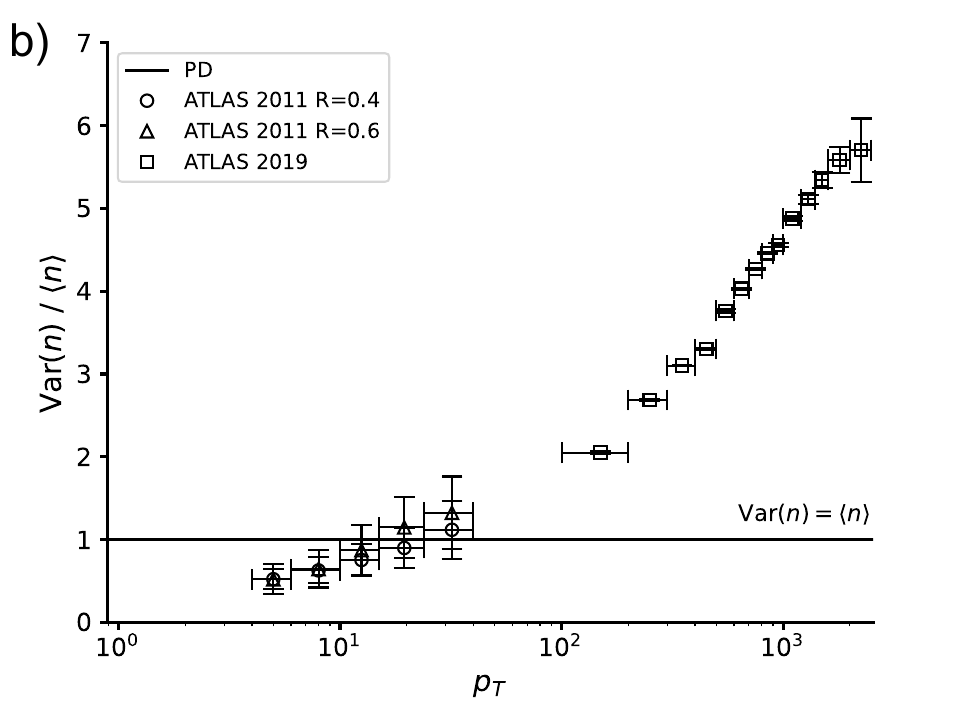}}
}
\end{tabular}
\caption{Variance as a function of  a) $\bar{n}$ and  b) $p_T$. Data are      
from Refs. \cite{atlas11} and \cite{atlas19}. The solid line shows the result 
obtained with the Poisson distribution.}
\label{fig2}
\end{figure}
\begin{figure}[!t]
\begin{tabular}{ccc}
\centerline{
{\includegraphics[height=4.5cm]{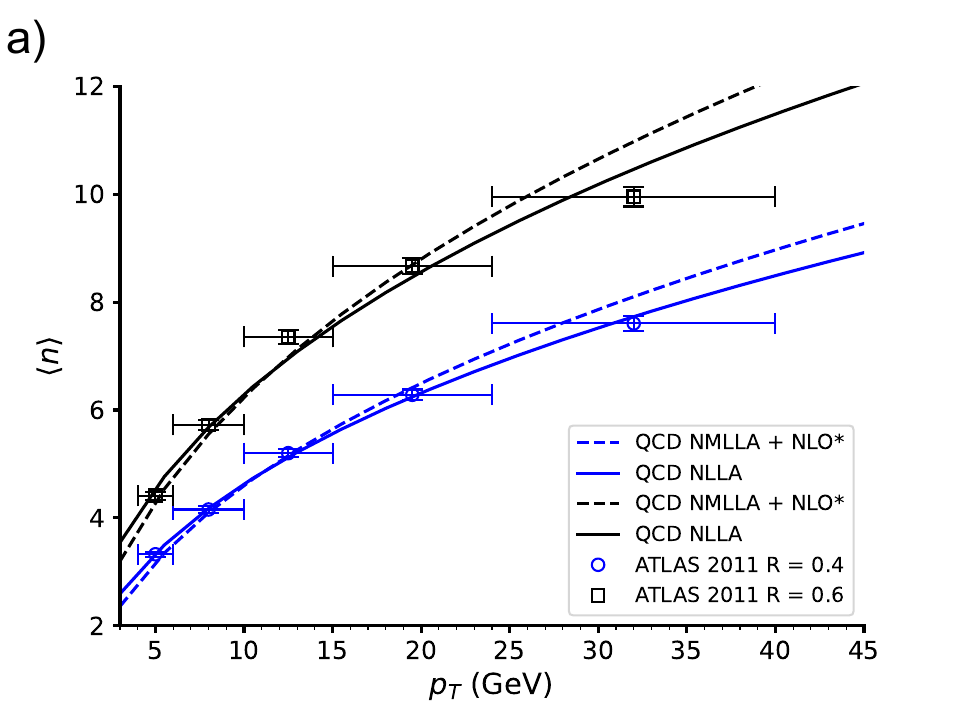}}
{\includegraphics[height=4.5cm]{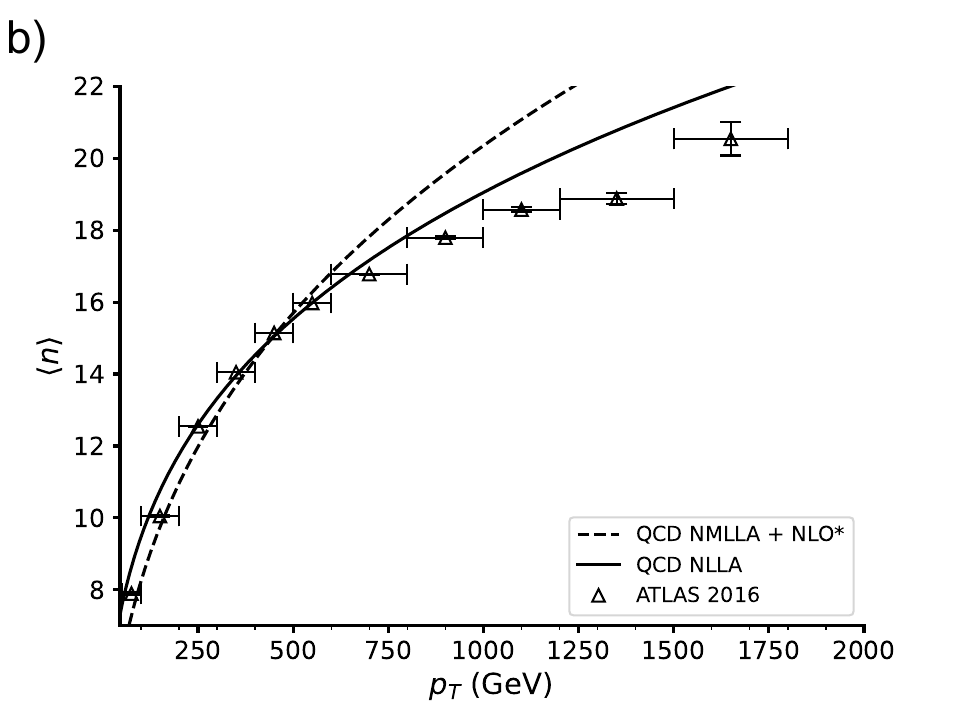}}
{\includegraphics[height=4.5cm]{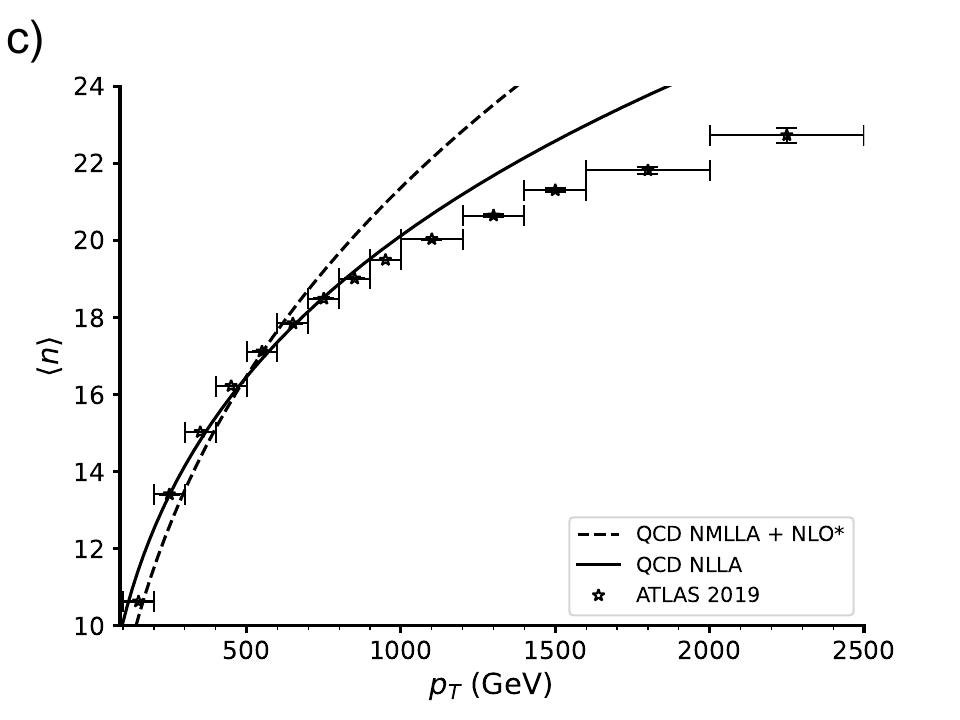}}
}
\end{tabular}
\caption{Average multiplicities (data points) obtained  from  ATLAS, in  a)  Ref.~\cite{atlas11}, b)  
Ref.~\cite{atlas16}  and  c)  Ref. \cite{atlas19}. The curves show the   
fits with the theoretical expressions Eq. (\ref{nlla}) (solid line) and Eq. (\ref{nmlla}) (dashed line).}
\label{fig3} 
\end{figure}  
In Fig.~\ref{fig1} we plot the ATLAS data on jet
multiplicities in the KNO form.  In Fig.~\ref{fig1}a and Fig.~\ref{fig1}b we show the data from 
Ref.~\cite{atlas11} on lower $p_T$ jets. The two sets refer to two values of the jet $R$ variable. 
We clearly see that the data violate  KNO scaling. However,  if we consider the higher $p_T$ jets measured 
in Ref.~\cite{atlas19} we observe the onset of scaling, as shown in  Fig.~\ref{fig1}c, specially at 
$p_T > 300$ GeV. This change of behavior can be more clearly seen if we plot the ratio between the variance and the   
average multiplicity ($Var(n)/\bar{n}$) which is equal to one for a Poisson distribution. In Fig.~\ref{fig2} we can see that this 
ratio is below one for lower $\bar{n}$ (Fig.~\ref{fig2}a) and for lower $p_T$ (Fig.~\ref{fig2}b) and 
around  $\bar{n} \simeq 10$  or $p_T \simeq 30$ GeV there is a clear change. The ratio becomes larger than one 
and we go from a sub-poissonian to a super-poissonian distribution. It is tempting to associate this broadening of the multiplicity 
distribution with the transition from quark to gluon initiated jets. In Ref. \cite{opal98} the properties of quark and gluon jets were 
studied. In particular, it was found that the dispersion D of the multiplicity distribution from jets was 
$D_g \simeq 4.37$ and $D_q \simeq 4.30$ for gluon and quark jets respectively. The errors quoted in  \cite{opal98} are very large but 
these numbers suggest that gluon jets are broader than quark ones and the onset of the dominance of the former could be the dynamical 
cause of the behavior observed in  Fig.~\ref{fig2}.  

From the low $p_T$ data \cite{atlas11} we have extracted the average multiplicities for the  $R=0.4$ and $R=0.6$ sets. For the 
higher $p_T$ sets the average multiplicities were already given in \cite{atlas19} and \cite{atlas16}.
From the theoretical point of view, the definition of the multiplicity in a jet can be rather tricky.  
The total hadronic multiplicity within a jet can be obtained from the jet fragmentation function and it was studied 
in perturbative QCD in several works \cite{mu81,we84,wema,dre94,drega,re14}. 
Very recently these calculations have been done with higher precision (see \cite{so22} for a recent review of the literature). 
Here, for simplicity, we shall use the  analytical formulas derived  from perturbative QCD in the next-to-leading-logarithmic  
approximation (NLLA) \cite{we84} and also, more recently, in  the  next-to-modified-leading-log approximation (NMLLA) including
next-to-leading-order (NLO) corrections to the $\alpha_s$ strong coupling \cite{re14}.  These  expressions were successfully applied to 
fit the multiplicities measured in $e^+ e^-$ collisions \cite{delphi98}. The NLLA expression is given by \cite{we84}:  
\beq
\bar{n}_{ch} = a \, [\alpha_s(p_T)]^b \, e^{c/\sqrt{\alpha_s(p_T)}} \,
[1 + d \, \sqrt{\alpha_s(p_T)}]
\label{nlla}
\eeq
where \cite{we84}:
$$
b = \frac{1}{4} + \frac{10}{27} \frac{N_f}{\beta_0} = 0.49 
\hskip1cm
c=  \frac{\sqrt{96 \pi}}{\beta_0} = 2.27
\hskip1cm
\alpha_s(Q^2) = \frac{4 \, \pi}{\beta_0 \, ln(Q^2/\Lambda^2)}
              - \frac{\beta_1 \, ln \, ln (Q^2/\Lambda^2)}{\beta_0^3 \, 
              ln^2 (Q^2/\Lambda^2)}
$$
with $\beta_0 = 11 - 2/3 N_f$, $\beta_1 = 102 - 38/3 N_f$ and  $\Lambda = 0.15$ GeV.  The NMLLA-NLO expression 
reads \cite{re14}: 
\beq
\bar{n}_{ch} = \mathcal{K}_{ch}  \, exp \, \left[ 2.50217 \, \sqrt{Y}  - 0.491546 \,  ln \, Y  - (0.06889 - 0.41151 \,  ln \,  Y ) 
\frac{1}{\sqrt{Y}}   +  (0.00068 - 0.161658 \,  ln \,  Y ) \frac{1}{Y} \right] \\
\\
\label{nmlla}
\eeq
where
$$ 
Y = ln \,(p_T/\Lambda_{QCD})
$$  
In the above expressions all the parameters have already been fixed so as to 
reproduce the multiplicity distributions measured in $e^+ e^-$  collisions at LEP and at energies ranging from $ 2 < \sqrt{s} < 200$ GeV. 
In (\ref{nlla}) the normalization $a$ and the (higher order corrections) parameter $d$ were adjusted. In (\ref{nmlla}) the 
normalization $\mathcal{K}_{ch}$ and $\Lambda_{QCD}$ were adjusted. Both expressions were able to yield very good fits. We 
assume that each of the two jets in $e^+ e^-$ collisions is initiated by one highly energetic parton, in the same way as the jets observed 
in $pp$ collisions. Therefore, apart from a normalization factor, the formulas (\ref{nlla}) and (\ref{nmlla}) can be applied to the 
average multiplicities studied in this work.

In Fig.~\ref{fig3} we show the average multiplicities. As it can be seen, the low $p_T$ data
from Ref.~\cite{atlas11} in Fig.~\ref{fig3}a  are well reproduced both by (\ref{nlla}) and (\ref{nmlla}).
The parameters obtained from the fits are shown in Table \ref{tab1}. In comparison to the jets measured in $e^+e^-$,  both the 
normalization factor and $\Lambda_{QCD}$ are systematically smaller. These expressions work well for the higher $p_T$ sets from Ref.~\cite{atlas16}  shown in Fig.~\ref{fig3}b  and also from Ref.~\cite{atlas19} shown in  Fig.~\ref{fig3}c up to 
$p_T \simeq 500 - 600$ GeV. Up to this point,  Eq. (\ref{nlla}) and Eq. (\ref{nmlla})    
seem to capture very well the energy dependence of the data.  Beyond this point, they overshoot the data. In this region it is possible 
that gluon recombination (not yet included in the calculations)  starts to play a role. In fact $gg \to g$  would reduce the
number of produced partons (and hadrons). Qualitatively this effect would go in the right direction to reproduce the data. 

\begin{table}[h]
\centering
    \begin{tabular}{|c|c|c|c|c|c|}
            \hline
                        & ATLAS 2011 R=0.4  & ATLAS 2011 R=0.6  & ATLAS 2016    & ATLAS 2019    & $e^+e^-$ \\
            \hline
            \hline
$\mathcal{K}_{ch}$                & 0.04(1)           & 0.06(3)           & 0.03(2)       & 0.03(1)       & 0.117(1) \\
            \hline
$\Lambda_{\mathrm{QCD}}$ (GeV)   & 0.15(8)           & 0.15(12)          & 0.15(22)      & 0.15(19)      & 0.191(13)\\
            \hline
$a$                     & 0.042(5)          & 0.05(2)           & -0.012(9)     & -0.015(7)     & 0.53(6)  \\
            \hline
$d$                     & 0.9(4)            & 1.1(1.4)          & -13(8)        & -11(4)        & 1.11(39)  \\
            \hline
    \end{tabular}
\caption{Fitted values of $\mathcal{K}_{ch}$ and $\Lambda_{\mathrm{QCD}}$  in Eq.(\ref{nmlla}) and values of $a$ and $d$ in Eq.(\ref{nlla})
for all data sets and the corresponding values of these parameters obtained from $e^+e^-$ collisions in \cite{re14} and 
\cite{delphi98}.}
        \label{tab1}
\end{table}

\section{The Sub-Poissonian distribution}

A Sub-Poissonian distribution (SPD) is a probability distribution that has a smaller variance than the Poisson one with the same mean. A  distribution which has a larger variance is called  Super-Poissonian  and to this class belongs also the widely used negative binomial distribution (NBD). In \cite{shi,PC-CCS} the SPD   was introduced in the context of particle physics and applied to study  MDs measured 
at lower energies. 

Since the SPD has not been used very often, it is worth saying  a few words about its origin and meaning. In particular, we would like to 
show how it can be obtained from a stochastic Markov process with multiplicity-dependent birth and death rates. 

Let  $P(n,t)$ be the probability of having $n$ particles at time $t$ and let us consider a very general birth-death process given by the 
following equations:
\begin{eqnarray}
\!\!\!\!\! \frac{d P (0,t)}{d t} &=& -\lambda_0 P(0,t) + \mu_1 P(1,t), \label{NZW-4a}\\
\!\!\!\!\! \frac{d P (n,t)}{d t}  &=& -\left( \lambda_n + \mu_n \right) P(n,t) + \lambda_{n-1} P(n-1,t) + \mu_{n+1} P(n+1,t), 
\label{NZW-4b}
\end{eqnarray}
where $\lambda_n$ and $\mu_n$ are the birth and death rates when the multiplicity is $n$. 
Let us further assume that:
\begin{equation}
\lambda_n = \lambda {(n+1)^{-\delta}}+ \sigma\quad{\rm and}\quad
\mu_n=n \mu, 
\label{NZW-5}
\end{equation}
Then, in the steady state, when $\frac{d P(n,t)}{d t}=0$, Eq. (\ref{NZW-4b}) yields: 
\begin{equation}
-\left[ \lambda(n+1)^{-\delta}+ \sigma + n \mu \right]P(n) + (\lambda n^{-\delta}+  \sigma)P(n-1) + (n+1) \mu P(n+1) = 0 
\label{NZW-6}
\end{equation} 
Introducing the notation
\begin{equation}
\frac{\lambda}{\mu} = \alpha  \quad{\rm and}\quad \frac{ \sigma}{\mu}
=\alpha_0, 
\label{NZW-7}
\end{equation}
Eq. (\ref{NZW-6}) can be rewritten as : 
\begin{equation}
- \left[ \alpha(n+1)^{-\delta} +\alpha_0 + n \right ]P(n) + \left [ \alpha n^{-\delta}+\alpha_0 \right ]P(n-1) + (n+1) P(n+1) = 0
\label{NZW-8}
\end{equation}
which yields the following recurrence relation
\begin{equation}
(n+1) P(n+1) = \left [ \alpha(n+1)^{-\delta} + \alpha_0  \right ]P(n) = g(n) \, P(n)  
\label{NZW-9}
\end{equation}
where $g(n)=(n+1)P(n+1)/P(n) = \alpha(n+1)^{-\delta} + \alpha_0$. Knowing $P(0)$, with the above expression we can construct the
multiplicity distribution:
\begin{equation}
P(n) = \frac{P(0)}{n!}\prod_{i=0}^{n-1} g(i) = \frac{P(0)}{n!}\prod_{i=0}^{n-1}  \left[ \alpha(i+1)^{-\delta} + \alpha_0 \right]  
\label{PN}
\end{equation}
Choosing $\alpha_0 >0$ and $\delta =-1$ we obtain the  Negative Binomial distribution
\begin{equation}
P(n) = \frac{\Gamma(n+k)}{\Gamma(n+1)\Gamma(k)}\alpha^n(1-\alpha)^k.
\label{NBD}
\end{equation}
where  $k=1+\alpha_0/\alpha$. 
In the particular case when  $\alpha_0 = 0$,
the parameter $k=1$ and  
the above expression reduces to the  geometric
(Bose-Einstein) distribution: 
\begin{equation}
P(n) = \alpha^n(1-\alpha). 
\label{GD-BE}
\end{equation} 
Setting  $\delta=0$ in (\ref{PN})  we obtain the  Poisson distribution :  
\begin{equation}
P(n) = \frac{\alpha^n}{n!} \exp(-\alpha).
\label{PD}
\end{equation}
For $\delta > 0$ we get the sub-Poissonian distribution:
\begin{equation}
P(n) = c\frac{\alpha^n}{(n!)^{1+\delta}},
\label{SP}
\end{equation}
where $c$ is a normalization factor. Notice that when $\delta = -1$ the above expression becomes (\ref{GD-BE}) apart from a constant factor.

\section{From Sub-Poisson to Negative Binomial Distribution}

\begin{figure}[!t]
\begin{tabular}{cc}
{\includegraphics[height=6.0cm]{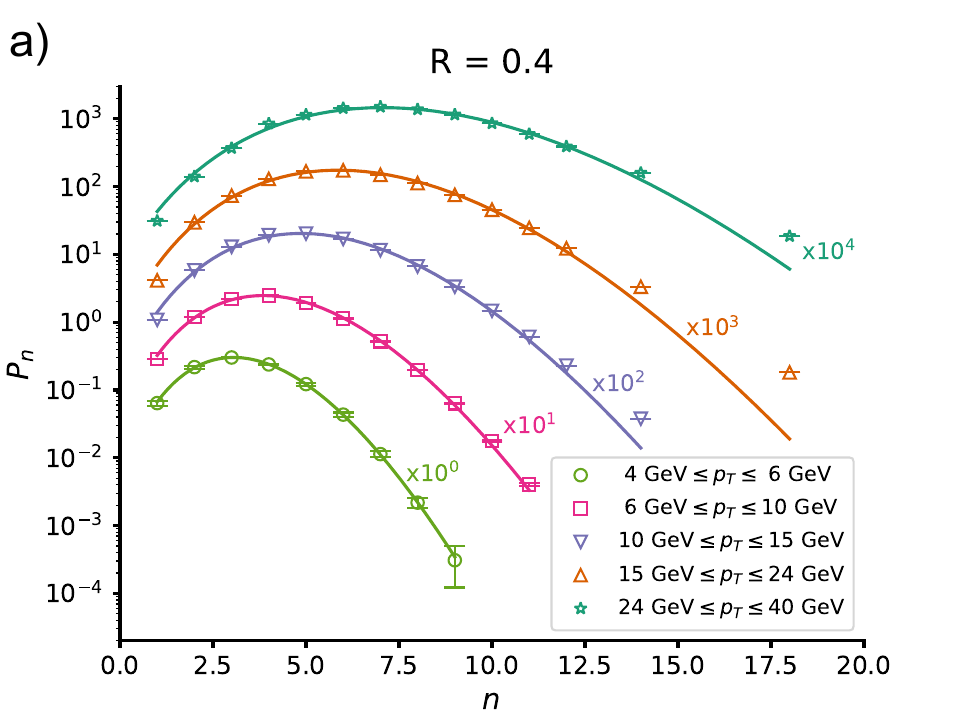}}
{\includegraphics[height=6.0cm]{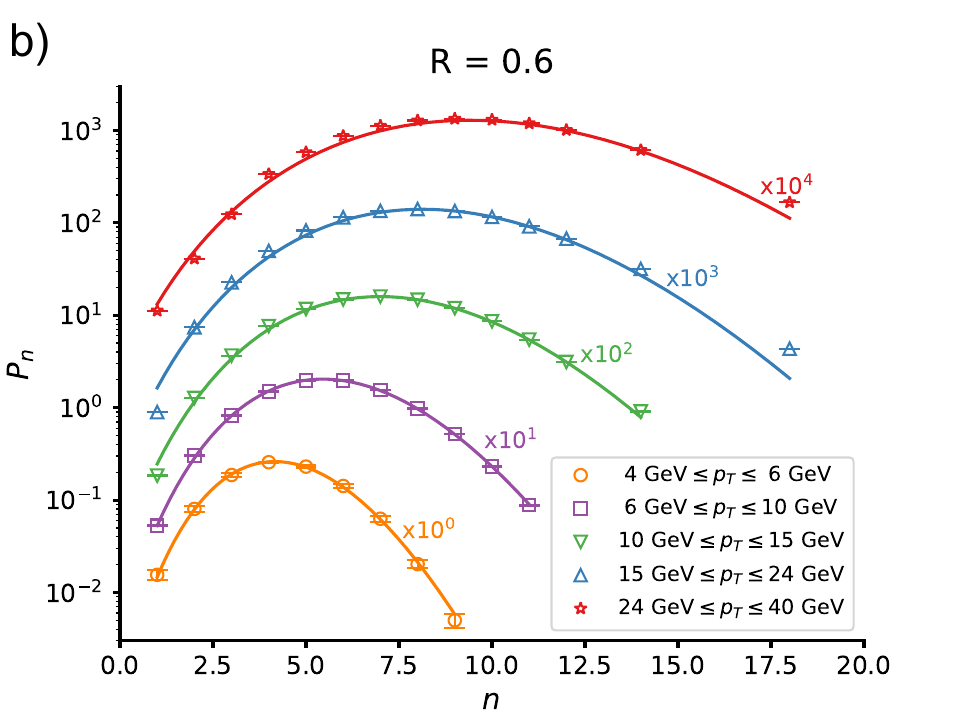}} \\
\\
{\includegraphics[height=6.0cm]{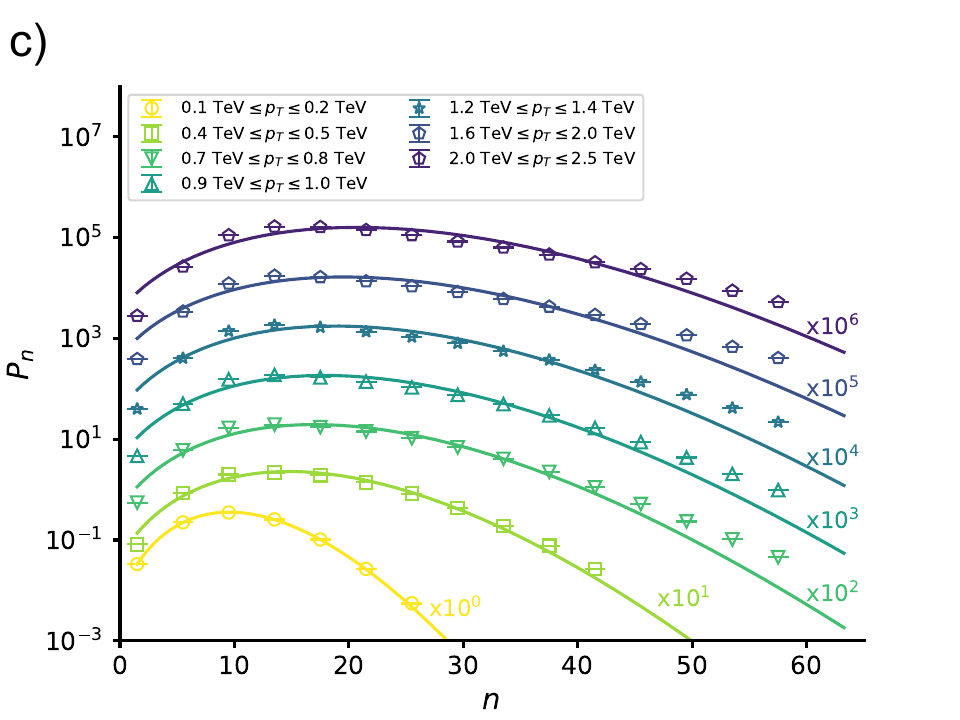}}
{\includegraphics[height=6.0cm]{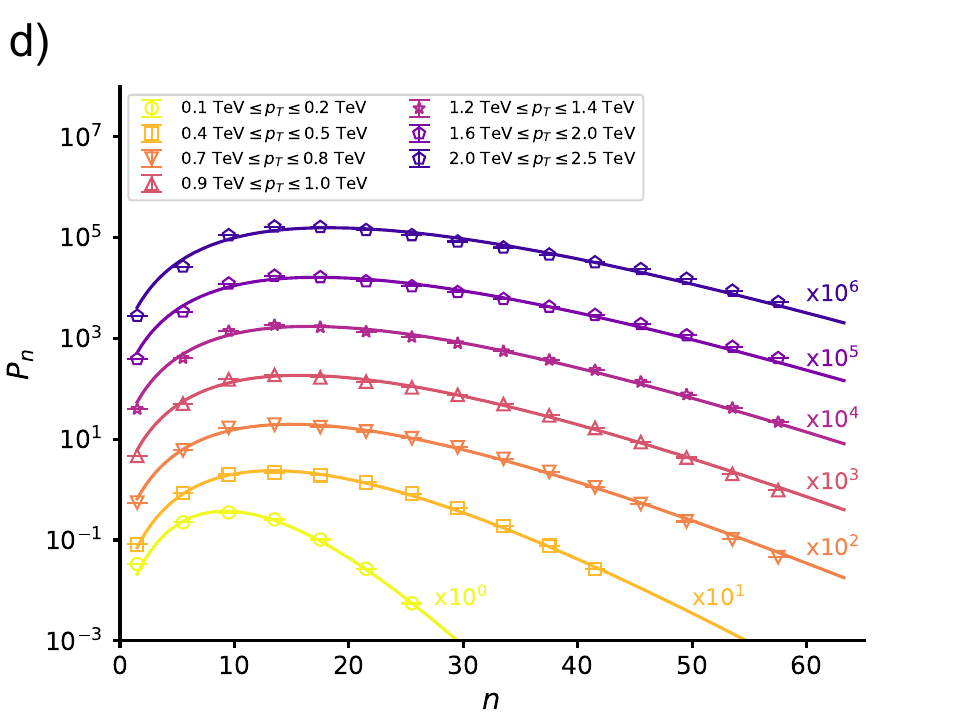}}
\end{tabular}
\caption{Fits of the ATLAS data on jet multiplicity distributions for  a) $R=0.4$~\cite{atlas11} , b) $R=0.6$ ~\cite{atlas11},  and data from Ref.~\cite{atlas19} fitted with a c) Sub-Poisson distribution (\ref{SP}) and  d) with a Negative Binomial distribution (\ref{NB}).}
\label{fig4}
\end{figure}

In this section we shall use the form (\ref{SP}) to study the ATLAS data on multiplicity distributions in jets. We will extend the work \cite{greg22} and  fit all the ATLAS data from Ref. \cite{atlas11} and also from Ref. \cite{atlas19}.     
We will fix $\alpha$ and $\delta$ adjusting  (\ref{SP}) to the data. The normalization factor $c$ is given in terms of $\alpha$ and 
$\delta$ as: 
\beq
c = \left( \sum_{1}^{N_{max}} \frac{\alpha ^n}{(n!)^{1+\delta}} 
\right)^{-1}
\label{alpha}
\eeq
where the $N_{max}$ is the number of data points. The results are shown in Fig.~\ref{fig4}a and  Fig.~\ref{fig4}b. 
As it can be seen, the SPD can reproduce very well the low $p_T$ ATLAS data. 
The fitted parameters $\alpha$ and $\delta$ are  shown in the Tables \ref{tab2} and \ref{tab3}, as well as the $\chi^2$ 
of the fits, which is always below 2.1. Because of the discrepancies in the large $n$ region, the $\chi^2$ of the high $p_T$ fits is 
unreasonably large and we do not show it in Table \ref{tab4}.  
As seen in Tables \ref{tab2}, \ref{tab3} and \ref{tab4}, for higher values of $p_T$ the $\delta$ parameter becomes negative, which signals the transition  from Sub-PD to Super-PD, best visible in Fig.~\ref{fig4}c. Notice that the simple formula (\ref{SP}) used here (with the parameter $\delta$ describing the departure from PD towards Sub-PD or Super-PD) approaches  the NBD limit (where $\delta = -1$). In 
Fig.~\ref{fig4}d we show lines obtained  with the NBD written in a form slightly different from 
(\ref{NBD}) and more convenient for our purposes. From (\ref{NBD}) it is easy to see that 
$\bar{n} = k \, \alpha/(1 - \alpha)$ and hence $\alpha^n \, (1 - \alpha)^k  = (\bar{n}^n  \, k^k)/( \bar{n} + k)^{(n + k)})$. Then (\ref{NBD}) 
can be rewritten as: 
\beq
P(n) = \frac{\Gamma(k+n)}{\Gamma(k) \, \Gamma(n+1)} \frac{\bar{n}^n  \, k^k}{(\bar{n}+k)^{n+k}}
\label{NB}.
\eeq  
The fits are very good. They show that $k$ decreases with the jet $p_T$ in the same way as it decreases with the energy in  NBD fits of the 
minimum bias multiplicity distributions \cite{ge22}. They also show that $\bar{n}/k$ increases with $p_T$ indicating the approach to 
KNO scaling, which is reached when $\bar{n} >> k$. Indeed, for KNO scaling, we expect the moments of the 
$P(z = n/\bar{n})$ distribution to be independent of $\bar{n}$. In the case of the NBD distribution, the second central moment of the $P(z)$ 
distribution is
\begin{equation}
Var(z) = \frac{Var(n)}{\bar{n}^2} = \frac{1}{\bar{n}} + \frac{1}{k},
\label{Var}
\end{equation}
For $\bar{n} \gg k$ we have an approximate scaling.  In fact, it was shown in \cite{dumi12}  that when this ratio reaches $~6$ one already 
has a very good scaling.  From the last entries of Table \ref{tab5} we have ratios close to this number. 
\begin{table}[h]
\centering
    \begin{minipage}[b]{0.48\hsize}\centering
        \begin{tabular}{|c|c|c|c|}
                        \hline
$p_T$ range (GeV) & $\alpha$ & $\delta$ & $\chi^2$ \\
            \hline
            \hline
[4,6]       & 12.8(2)   & 1.00(1)  & 0.06     \\
            \hline
[6,10]      & 11.1(4)   & 0.64(2)  & 0.32     \\
            \hline
[10,15]     & 10.2(5)   & 0.38(3)  & 1.36     \\
            \hline
[15,24]     & 9.2(6)    & 0.20(4)  & 2.08     \\
            \hline
[24,40]     & 7.2(5)    & -0.02(4) & 1.13     \\
                        \hline
                \end{tabular}
\caption{Fitted $\alpha$ and $\delta$ from Eq. (\ref{SP}) for R=0.4 and data from \cite{atlas11}.}
        \label{tab2}
    \end{minipage}
    \hfill
    \begin{minipage}[b]{0.48\hsize}\centering
        \begin{tabular}{|c|c|c|c|}
        \hline
$p_T$ range (GeV) & $\alpha$ & $\delta$ & $\chi^2$ \\
            \hline
            \hline
            \hline
[4,6]       & 21.4(9)   & 0.98(3)  & 0.30     \\
            \hline 
[6,10]      & 15.7(3)   & 0.54(1)  & 0.06     \\
            \hline 
[10,15]     & 10.6(4)   & 0.17(2)  & 0.46     \\
            \hline 
[15,24]     & 8.5(6)    & 0.00(3)  & 1.55     \\
            \hline 
[24,40]     & 6.6(6)    & -0.18(4) & 1.32     \\
                        \hline
                \end{tabular}
        \caption{Fitted $\alpha$ and $\delta$ from Eq. (\ref{SP}) for R=0.6 and data from \cite{atlas11}.}
        \label{tab3}
    \end{minipage}
\end{table}
\begin{table}[h]
\centering
    \begin{minipage}[b]{0.48\hsize}\centering
        \begin{tabular}{|c|c|c|}
        \hline
$p_T$ range (GeV) & $\alpha$ & $\delta$ \\
            \hline
            \hline
            \hline
[100,200]       & 3.13(6)   & -0.505(8)  \\
            \hline 
[400,500]       & 2.30(9)   & -0.69(1)  \\
            \hline 
[700,800]       & 2.0(1)    & -0.74(2)  \\
            \hline 
[900,1000]      & 2.0(1)    & -0.76(2)  \\
            \hline 
[1200,1400]     & 1.9(1)    & -0.77(2) \\
            \hline 
[1600,2000]     & 1.9(1)    & -0.79(2) \\
            \hline 
[2000,2500]     & 1.9(2)    & -0.79(2) \\
                        \hline
                \end{tabular}
        \caption{Fitted $\alpha$ and $\delta$ from Eq. (\ref{SP}), for ATLAS 2019 data \cite{atlas19}.} 
        \label{tab4}
    \end{minipage}
    \hfill
    \begin{minipage}[b]{0.48\hsize}\centering
        \begin{tabular}{|c|c|c|}
        \hline
$p_T$ range (GeV) & $\bar{n}$ & $k$ \\
            \hline
            \hline
            \hline
[100,200]       & 10.5(1)   & 12(1)  \\
            \hline 
[400,500]       & 16.0(2)   & 7.4(4) \\
            \hline 
[700,800]       & 18.5(2)   & 6.0(2) \\
            \hline 
[900,1000]      & 19.6(2)   & 5.7(2) \\
            \hline 
[1200,1400]     & 20.8(3)   & 5.4(3) \\
            \hline 
[1600,2000]     & 22.0(4)   & 5.2(3) \\
            \hline 
[2000,2500]     & 23.0(4)   & 5.2(3) \\
                        \hline
                \end{tabular}
        \caption{Fitted $\bar{n}$ and $k$ from eq. (\ref{NB}), for ATLAS 2019 data \cite{atlas19}.} 
        \label{tab5}
    \end{minipage}
\end{table}

To summarize, we observe a transition from the low $p_T$ 
region, where there is no KNO scaling, to the high $p_T$ region, where we find the scaling shown in Fig.~\ref{fig1}c.  In our description 
this is related to the fact that the SPD given by (\ref{SP}) turns into a Super-Poisson distribution for large $p_T$ because  $\delta$ 
becomes negative. In turn, the Super-PD distribution transforms for $\delta = -1$ into  the geometric distribution 
(\ref{GD-BE}), i.e. the NBD  with $k=1$, for which  we have KNO scaling. Therefore, what we observe represents a gradual change in dynamics causing the gradual (with increasing $p_T$) emergence of KNO scaling. Multiplicity distributions measured at higher energies are, 
as shown in \cite{vertesi21,vertesi22}, better described by NBD. This fact is interesting because it 
means a transition from one dynamical regime to another \cite{shi,PC-CCS}.

A detailed interpretation of these results in terms of the QCD dynamics is beyond the scope of this work but there 
are hints which may help theorists. We now present a few  heuristic remarks, but further investigation is required for definite conclusions.
They are the following:\\

\vspace{0.3cm}

$(i)$ {\it  Parton Saturation}

\vspace{0.3cm}

The decreasing values of $\alpha$ suggest that the death rate must increase with the energy. In a parton cascade this would mean that the process $gg \to g$ becomes more important and when this happens we are approaching the gluon saturation regime. From 
Fig. \ref{fig3}  we see that the deviations from the standard  perturbative QCD calculations occur at $~1$ TeV and this energy is high enough 
for saturation effects to become visible.\\

$(ii)$ {\it Quark and gluon jets and phase space}

\vspace{0.3cm}

The initially positive values of $\delta$  render $P(n)$ narrow. This may be a consequence of phase space restrictions. At lower energies we have a smaller number of particles and energy-momentum conservation prevents large fluctuations. At higher energies and larger number of particles the fluctuations are also larger and $P(n)$ becomes broader. Alternatively, the narrowness of $P(n)$ at lower values of $p_T$  may indicate that the jets are initiated by quarks. A broader $P(n)$ would indicate the dominance of gluon initiated jets.\\

$(iii)$ {\it Threshold effects} 

\vspace{0.3cm}

Based on the "parton liberation” picture \cite{AHM} and on the ”local parton-hadron duality” \cite{DKTM} we expect the  
$P(n)$ obtained from the parton cascade to be similar to the $P(n)$ of observed hadrons. In the parton model the average number of partons is related to the deep inelastic structure function and at high energies it is assumed that $ \bar{n} \,  \propto \, G(x, Q^2)$, where $G$ is the gluon distribution. 
An increase in the number of gluons implies that the total momentum will be partitioned among more gluons and it becomes more likely to find a gluon with a small fraction of the total momentum. Particle production in the fusion of
two gluons with momenta $x_1$ and $x_2$ introduces a threshold $x_1 \, x_2 \geq m^2/s$, where $m$ is the (sum of the) mass(es) of the produced  hadron(s) and $\sqrt{s}$ is the proton-proton collision energy. Imposing this restriction will exclude  the lower $x$ domain of $G(x)$ where the number of gluons grows rapidly and thus will exclude the 
larger $n$ configurations. In other words, the  imposition of particle production in a collision changes the solution of the parton cascade 
equation, favoring a narrower $P(n)$ with smaller $\bar{n}$. This motivates 
the use of the power index $\delta$ in the birth rate in Eqs. (\ref{NZW-5}) and in (\ref{SP}). As the energy $\sqrt{s}$ increases the threshold constraint becomes less restrictive and allows for a larger number of gluons producing a larger number of  final hadrons. As a consequence $P(n)$ becomes 
broader and with a larger $\bar{n}$.  In Eq. (\ref{SP})  this behavior translates into a decreasing value of $\delta$.\\

$(iv)$ {\it Convolution, substructures and scaling} 

\vspace{0.3cm}

In minimum bias proton-proton collisions typically half of the energy is released in the central rapidity region and the other
half is carried by the remnants of the incoming protons. The fraction of the energy released in the central region may change with 
the energy \cite{igm}. This picture would suggest that the observed hadrons in the final state come from three sources. However the number of sources may be larger because the "central fireball" may be composed of subsystems, smaller fireballs. This depends on details of the dynamics,
i.e., perturbative or non-perturbative, with more or less string formation and decay, with or without thermalization, etc. In any case 
an important part of our understanding of the collision  is to characterize the sources. 

The number of sources will follow a distribution 
$F$ and each source will emit a number of hadrons with a distribution $G$. Therefore the final multiplicity distribution $P$ will be the 
result of the convolution $ P = F \otimes G$. Alternatively, we can fix the number of sources to be one (which might be appropriate for 
jets) and let the average number of hadrons produced from this source, $\bar{n}$, to fluctuate according to a distribution $F$. Along this 
line, in Ref. \cite{PB} (see also \cite{SM}) it was shown that, P will follow KNO scaling (at large n and for finite $n/\bar{n}$) 
if it can be written as a convolution where G is a Poisson distribution and  $F$ a gamma function. In this case, the convolution $ F \otimes G$ 
yielded a negative binomial distribution.

At lower energies, we do not observe scaling. On the other hand, at higher jet energies we observe empirically 
that KNO scaling is reached  and also that the negative binomial distribution reproduces the data very well. This suggests that the  limiting $P(n)$ for jets can be written as the convolution used in \cite{PB}, allowing us to make conjectures about the fluctuations of $\bar{n}$ and 
about the nature of the sources.

\vspace{0.5cm}
 
\section{Conclusions}

In this work we have studied the multiplicity distribution of charged particles within jets in proton-proton collisions, which were measured 
by the ATLAS collaboration. In the region $p_T < 500$ GeV  the mean multiplicity as a function of the jet transverse momentum 
is well fitted by the QCD-NLLA and QCD-NMLLA formulas. At higher values of $p_T$ these formulas overshoot the data. The low $p_T$ data \cite{atlas11} do not show KNO scaling, whereas the higher $p_T$ data \cite{atlas19}  gradually approach KNO scaling. 
This is in line with the PYTHIA simulations presented in Ref.~\cite{vertesi21}. 
Using   Eq. (\ref{SP}),  which, depending on the sign of the $\delta$ parameter describes Sub-Poissonian, Poissonian and Super-Poissonian distributions, we have fitted all the  existing data. However, at the highest values of $p_T$ the best fit is obtained with the negative binomial distribution. The ratio $\bar{n}/k$ of the NBD fits is large, giving quantitative support to the approach to KNO scaling. 
The results presented here illustrate the research potential of the analysis of multiplicity distributions in high-energy jets 
for various values of transverse momenta $p_T$. As our analysis shows, different $p_T$ ranges are described by  different  dynamics.

\vspace{0.5cm}

\noindent{\bf Acknowledgments:}

We are grateful to the brazilian funding agencies FAPESP, CNPq and CAPES and also to the INCT-FNA. GW was supported in part by the Polish Ministry of Education and Science, grant Nr 2022/WK/01.

\end{document}